\documentclass[aps,pra,twocolumn,groupedaddress,showpacs]{revtex4}
\usepackage{graphicx}
\usepackage{dcolumn}
\usepackage{bm}
\begin{document}

\title{Evidence for Bohmian velocities from conditional Schr\"{o}dinger equation}

\author{Hrvoje Nikoli\'c}
\affiliation{Theoretical Physics Division, Rudjer Bo\v{s}kovi\'{c}
Institute,
P.O.B. 180, HR-10002 Zagreb, Croatia.}
\email{hrvoje@thphys.irb.hr}

\date{\today}

\begin{abstract}
It is often argued that measurable predictions of Bohmian mechanics 
cannot be distinguished from those of a theory with arbitrarily modified  
particle velocities satisfying the same equivariance equation.
By considering the wave function of a closed system in a state 
with definite total energy, we argue that a distinction in measurable predictions 
is possible. Even though such a wave function is time-independent,
the conditional wave function for a subsystem depends on time through 
the time-dependent particle trajectories not belonging to the subsystem. 
If these trajectories can be approximated by classical trajectories, then
the conditional wave function can be approximated by a wave function 
which satisfies Schr\"{o}dinger equation in a classical time-dependent potential,
which is in good agreement with observations. However, such an approximation
cannot be justified for particle velocities significantly deviating from the
Bohmian ones, implying that Bohmian velocities are observationally preferred. 
\end{abstract}

\pacs{03.65.Ta, 03.65.Sq}

\maketitle

\section{Introduction}

The Bohmian formulation of quantum mechanics (QM) in terms 
of deterministic particle trajectories \cite{bohm},
originally developed to offer a possible resolution of the interpretative difficulties
with QM (see also \cite{book-bohm,book-hol,book-durr}),
today offers also a practical tool in various applications \cite{wyatt-prl,wyatt-book,oriols,apl-book}.
A recent work has demonstrated that Bohmian particle trajectories can even be weakly measured
\cite{weak1,weak2,weak-science,weak4}, so in a sense they
are no longer ``hidden variables''. Namely, these weak measurements
demonstrate that Bohmian trajectories are not more hidden
than the wave function, i.e., that both can be measured as properties 
of large statistical ensembles, but not as properties of individual systems.
Whether the Bohmian trajectories, or wave functions, or both, have reality at the level of individual systems
is observationally still unclear.

A frequent argument (see, e.g., \cite{weak1,weak2}) against reality of Bohmian trajectories is the claim that 
their measurable predictions cannot be distinguished from those of a theory with arbitrarily modified  
particle velocities satisfying the same equivariance equation
\begin{equation}\label{int1}
 \partial_t |\Psi|^2 +\sum_{a=1}^N \mbox{\boldmath $\nabla$}_a (|\Psi|^2 {\bf u}_a)=0 ,
\end{equation}
where $\Psi({\bf x}_1,\ldots,{\bf x}_N,t)$ is the wave function for $N$ particle
positions ${\bf x}_a$, and ${\bf u}_a({\bf x}_1,\ldots,{\bf x}_N,t)$ is the velocity
of particle $a$ as a function of time $t$ and all particle positions ${\bf x}_1,\ldots,{\bf x}_N$.
Eq.~(\ref{int1}) is satisfied for any velocity of the form 
\begin{equation}\label{int2}
{\bf u}_a={\bf v}_a +\frac{{\bf j}_a}{|\Psi|^2} ,
\end{equation}
where 
${\bf j}_a({\bf x}_1,\ldots,{\bf x}_N,t)$ is an {\em arbitrary} vector field
with the property
$ \sum_a \mbox{\boldmath $\nabla$}_a {\bf j}_a =0$,
${\bf v}_a$ is the Bohmian velocity 
$ {\bf v}_a=m_a^{-1}\mbox{\boldmath $\nabla$}_a S $,
$S({\bf x}_1,\ldots,{\bf x}_N,t)$ is defined by the wave function in the
polar form $\Psi=Re^{iS/\hbar}$, and $m_a$ is the mass of particle $a$. 
The arguments for preferring Bohmian velocity ${\bf u}_a={\bf v}_a$ (i.e., ${\bf j}_a=0$) 
include the analogy with classical Hamilton-Jacobi mechanics 
\cite{bohm,book-bohm,book-hol}, Galilean invariance \cite{book-durr}, and
the relation with weak measurement of velocity \cite{weak1,weak2}.  
Nevertheless, from an observational point of view neither of these arguments seems conclusive.

In this paper we present a new observationally supported argument for preferring
Bohmian velocity ${\bf u}_a={\bf v}_a$. For that purpose, suppose that 
a closed system (e.g., the whole Universe) is in a state with definite total energy $E$.
The Schr\"odinger equation implies that the wave function has the form
$\Psi({\bf x}_1,\ldots,{\bf x}_N,t)=\psi({\bf x}_1,\ldots,{\bf x}_N)e^{-iEt/\hbar}$, 
implying that the wave function is essentially time-independent, in the sense that
$|\Psi|^2=|\psi|^2$ does not depend on time. On the other hand, the observed world
clearly depends on time. How the observed time dependence can be reconciled with a time-independent 
wave function of the Universe $\psi$?

An obvious answer is that our Universe is simply not in a state with definite total energy,
but in a superposition of many different Hamiltonian eigenstates. However, such an answer
does not seem satisfying \cite{page} because such a state of matter in a superposition of different energies 
would imply that the gravitational field, determined by energy of matter, is in a superposition of 
states corresponding to macroscopically different gravitational fields, which would contradict
the classical appearance of observed gravitational fields. Just as decoherence theory
applied to coupling between charge and electromagnetic field explains the charge
superselection rule \cite{giulini,decoh1}, a similar superselection rule preventing superpositions
of different matter energies is expected from decoherence theory
applied to coupling between matter and gravity. 
This argument against time dependence in QM is also closely related to the problem of time in
canonical quantum gravity \cite{kuchar,isham}, where the total energy of matter and
gravity must be exactly zero at any point in space, thus preventing any time dependence
of the wave function.

A key to the resolution of the problem of time-dependence lies in the observation
that most interpretations of QM introduce some additional time dependence in the
system (e.g., through wave-function collapse, classical macro-world, 
free human interventions, or time-dependent hidden variables), 
implying that time evolution of the Universe is not described 
completely by the Schr\"{o}dinger equation alone. 
In Appendix \ref{SEC-interp} we briefly review how various
interpretations deal with it. The main purpose of the present paper is to explain
in detail how the Bohmian formulation of QM explains the origin of time
dependence in closed quantum systems with definite energy, and 
how that provides an observational evidence for preferring
Bohmian velocities over other velocities in (\ref{int2}) consistent with (\ref{int1}).

The basic idea is not difficult to understand. Suppose that one is interested
in a subsystem consisting of $n<N$ particles. For that purpose one can consider
the {\em conditional wave function}
\begin{equation}\label{int5}
 \psi_c({\bf x}_1,\ldots,{\bf x}_n,t) =\psi({\bf x}_1,\ldots,{\bf x}_n,{\bf X}_{n+1}(t),\ldots,{\bf X}_N(t)) , 
\end{equation}
where ${\bf X}_{n+1}(t),\ldots,{\bf X}_N(t)$ are the actual particle trajectories of all other
particles, with velocities $d{\bf X}_a(t)/dt ={\bf u}_a$.  
The conditional wave function has a well-understood role in explaining the illusion of
wave-function collapse \cite{durr-big,book-durr}. In this paper we study a different role of
conditional wave function, as a tool to understand the origin of time-dependence itself.
If the total wave function $\psi({\bf x}_1,\ldots,{\bf x}_N)$ satisfies the time-independent
Schr\"odinger equation
\begin{equation}\label{int6}
\left[ \sum_{a=1}^N  \frac{-\hbar^2}{2m_a}\mbox{\boldmath $\nabla$}_a^2
+V({\bf x}_1,\ldots,{\bf x}_N) \right] \psi = E\psi ,
\end{equation}
then, under certain approximations, we will find out that the conditional wave function satisfies
\begin{equation}\label{int7}
 \left[ \sum_{a=1}^n  \frac{-\hbar^2}{2m_a}\mbox{\boldmath $\nabla$}_a^2
+V_c({\bf x}_1,\ldots,{\bf x}_n,t) \right] \psi_c = i\hbar\partial_t \psi_c ,
\end{equation}
where 
\begin{equation}\label{Vcgen}
V_c({\bf x}_1,\ldots,{\bf x}_n,t)=V({\bf x}_1,\ldots,{\bf x}_n,{\bf X}_{n+1}(t),\ldots,{\bf X}_N(t)) .
\end{equation}
If ${\bf X}_{n+1}(t),\ldots,{\bf X}_N(t)$ are {\em classical} trajectories, then
(\ref{int7}) can be viewed as a quantum system of $n$ particles in a 
{\em classical background time-dependent potential}, where the time-dependence of the potential
comes from classical time-dependence of the environment.
Schr\"odinger equation in a classical time-dependent potential is a frequently used approximation
known to be in good agreement with observations. This means that (\ref{int7}) agrees with observations
only if the trajectories ${\bf X}_{n+1}(t),\ldots,{\bf X}_N(t)$ {\em can} be approximated
by the classical ones. On the other hand, 
we shall see that such an approximation is justified if ${\bf j}_a=0$
in (\ref{int2}), but is not justified for arbitrary (sufficiently large) ${\bf j}_a$ compatible with (\ref{int1}).
Therefore, observations prefer the Bohmian velocity ${\bf u}_a={\bf v}_a$. 
(Even though, of course, no observation can rule out sufficiently small ${\bf j}_a$.)

The rest of the paper is organized as follows. For the sake of better understanding of 
conditional Schr\"odinger equation, in Sec.~\ref{SEC-clas} we first develop
conditional {\em classical} mechanics, with emphasis on the concepts of
conditional Hamiltonian and conditional Hamilton-Jacobi equation.
Then in the central section, Sec.~\ref{SEC-quant}, we generalize these results in classical
mechanics to those in quantum mechanics, by presenting a detailed derivation of the
conditional Schr\"odinger equation. In Sec.~\ref{SEC-phys} we discuss
the physical relevance of the conditional Schr\"odinger equation, where we explain
in more detail how conditional Schr\"odinger equation explains the origin of time evolution in 
closed quantum systems with definite energy, and how that prefers
Bohmian velocity over other velocities consistent with (\ref{int1}). 
Finally, the conclusions are drawn in Sec.~\ref{SEC-concl}.

\section{Conditional classical mechanics}
\label{SEC-clas}

\subsection{Conditional Hamiltonian}
\label{SEC3.1}

Consider a closed classical system of two particles with positions $x_1$ and $x_2$, 
described by the Hamiltonian
\begin{equation}\label{c1}
H(x_1,p_1,x_2,p_2)=\frac{p_1^2}{2m_1}+\frac{p_2^2}{2m_2}+V(x_1,x_2) . 
\end{equation}
(We consider only two particles moving in only one dimension just to simplify the notation.
The generalization of our results to arbitrary number of particles in three dimensions
is trivial.)
We distinguish the abstract time-independent canonical variables $x_a$, $p_a$, $a=1,2$, from the
physical time-dependent particle trajectories $X_a\equiv X_a(t)$, $P_a\equiv P_a(t)$.
The latter are given by the Hamilton equations of motion
\begin{eqnarray}\label{c1.1}
 \frac{dX_a}{dt}=\frac{\partial H(X_1,P_1,X_2,P_2)}{\partial P_a}, 
\nonumber \\
\frac{dP_a}{dt}=-\frac{\partial H(X_1,P_1,X_2,P_2)}{\partial X_a} .
\end{eqnarray}

Now assume that trajectory $X_2(t)$, $P_2(t)$ of the second particle is known.
The dynamics of the first particle can be described by the {\em conditional Hamiltonian}
\begin{equation}\label{c1.2}
 H_c(x_1,p_1,t)=H(x_1,p_1,X_2(t),P_2(t) ) .
\end{equation}
Eq.~(\ref{c1.1}) for $a=1$ is equivalent to 
\begin{equation}\label{c1.3}
 \frac{dX_1}{dt}=\frac{\partial H_c(X_1,P_1,t)}{\partial P_1}, 
\;\;\;\;
\frac{dP_1}{dt}=-\frac{\partial H_c(X_1,P_1,t)}{\partial X_1}, 
\end{equation}
which are nothing but Hamilton equations of motion for the system described by the
conditional Hamiltonian. 
Using (\ref{c1}), we see that (\ref{c1.2}) can be written in a more explicit form
\begin{equation}\label{c1.4}
H_c(x_1,p_1,t)=\frac{p_1^2}{2m_1}+\frac{P_2^2(t)}{2m_2}+V_c(x_1,t) , 
\end{equation}
where
\begin{equation}
 V_c(x_1,t)=V(x_1,X_2(t)) .
\end{equation}
The term $P_2^2(t)/2m_2$ in (\ref{c1.4})
depends only on time, not on $x_1$ or $p_1$,
so this term is irrelevant for the Hamilton equations of motion (\ref{c1.3}).
Consequently, (\ref{c1.3}) is valid also if $H_c$ given by (\ref{c1.4}) is replaced by
a simpler conditional Hamiltonian
\begin{equation}\label{c1.5}
 \tilde{H}_c(x_1,p_1,t)=\frac{p_1^2}{2m_1}+V_c(x_1,t) .
\end{equation}
Unlike (\ref{c1}), the conditional Hamiltonian (\ref{c1.5}) has an explicit dependence on time
and the corresponding energy is not conserved. This is because only the first particle 
is a dynamical variable in the conditional Hamiltonian, and consequently the system described
by the conditional Hamiltonian is not closed.

\subsection{Conditional Hamilton-Jacobi equation}
\label{SEC3.2}

The conditional Hamiltonian can be used to introduce the conditional 
Hamilton-Jacobi equation. We start from the total principal function
$S_{\rm tot}(x_1,x_2,t)$ satisfying the Hamilton-Jacobi equation for the
whole system
\begin{equation}\label{c1.6}
H(x_1,\partial_1 S_{\rm tot},x_2,\partial_2 S_{\rm tot})=-\partial_t S_{\rm tot},  
\end{equation}
where $\partial_a \equiv \partial / \partial x_a$, $\partial_t \equiv \partial / \partial t$.
Since the total energy $E$ is conserved, the solution of (\ref{c1.6}) has the form
\begin{equation}\label{c2}
 S_{\rm tot}(x_1,x_2,t)=S(x_1,x_2)-Et,
\end{equation}
so (\ref{c1.6}) reduces to the time-independent Hamilton-Jacobi equation
\begin{equation}\label{c3}
H(x_1,\partial_1 S,x_2,\partial_2 S)=E .  
\end{equation}

To simplify further mathematical expressions, we introduce a short-hand notation 
\begin{eqnarray}
& F(x_1,x_2) \equiv F , &
\nonumber \\
& F(x_1,X_2(t)) \equiv F_c , &
\nonumber \\
& F(X_1(t),X_2(t)) \equiv F_t , &
\end{eqnarray}
for any function $F$.
With this notation, the particle trajectories (\ref{c1.1}) with Hamiltonian (\ref{c1})
are equivalent to
\begin{equation}\label{c4}
 \frac{dX_a(t)}{dt}=\frac{1}{m_a}(\partial_a S)_t ,
\end{equation}
where
\begin{equation}
m_a \frac{dX_a(t)}{dt} = P_a(t).
\end{equation}

Now assume that $X_2(t)$ is known. 
Clearly, (\ref{c4}) implies that the remaining trajectory $X_1(t)$ can be calculated as
\begin{equation}\label{c6}
\frac{dX_1(t)}{dt}=\frac{1}{m_1}(\partial_1 S_c)_t ,
\end{equation}
where
\begin{equation}\label{c5}
 S_c(x_1,t)=S(x_1,X_2(t)) 
\end{equation}
is the conditional principal function.

On the other hand, we have seen in Sec.~\ref{SEC3.1} that
dynamics of the remaining particle can equivalently be described
by the simplified conditional Hamiltonian (\ref{c1.5}).
Since Hamilton formalism is equivalent to
Hamilton-Jacobi formalism, it implies that the remaining particle can be described
by the {\em conditional Hamilton-Jacobi equation}
\begin{equation}\label{c8}
\tilde{H}_c(x_1,\partial_1 \tilde{S}_c,t)=-\partial_t \tilde{S}_c ,
\end{equation}
the solution of which determines the particle trajectory
\begin{equation}\label{c6.1}
\frac{dX_1(t)}{dt}=\frac{1}{m_1}(\partial_1 \tilde{S}_c)_t .
\end{equation}
Here $\tilde{S}_c(x_1,t)$ is a solution of (\ref{c8}) which gives the same velocity
$dX_1/dt$ (at any given $x_1$) as $S_c(x_1,t)$ given by (\ref{c5}), so
\begin{equation}\label{c9}
 \partial_1 \tilde{S}_c = \partial_1 S_c 
\end{equation}
for any $x_1$. Consequently, $\tilde{S}_c$ and $S_c$ are related as
\begin{equation}\label{c10}
 \tilde{S}_c(x_1,t)=S_c(x_1,t) +f(t) ,
\end{equation}
where $f(t)$ is some function which depends only on $t$. 

\section{Conditional quantum mechanics}
\label{SEC-quant}

\subsection{Conditional wave function in the polar form}
\label{SEC-polar}

Now we are ready to generalize the results of Sec.~\ref{SEC-clas} to quantum mechanics.
The total system is described by the total wave function $\psi_{\rm tot}(x_1,x_2,t)$
satisfying the Schr\"{o}dinger equation
\begin{equation}\label{q11pr}
H(x_1,\hat{p}_1,x_2,\hat{p}_2) \psi_{\rm tot} = i\hbar \partial_t \psi_{\rm tot} ,
\end{equation}
where
\begin{equation}
\hat{p}_a=-i\hbar \partial_a . 
\end{equation}
We are interested in a solution with a definite total energy $E$, so 
$\psi_{\rm tot}(x_1,x_2,t)=\psi(x_1,x_2)e^{-iEt/\hbar}$ and
(\ref{q11pr}) reduces to
\begin{equation}\label{q11}
H(x_1,\hat{p}_1,x_2,\hat{p}_2) \psi = E \psi .
\end{equation} 
The solution can be written in the polar form
\begin{equation}\label{q12}
 \psi(x_1,x_2)=R(x_1,x_2)e^{iS(x_1,x_2)/\hbar} ,
\end{equation}
so complex equation (\ref{q11}) is equivalent to a set of two real equations
\begin{equation}\label{q13}
 H_Q(x_1,\partial_1 S,x_2,\partial_2 S)=E ,
\end{equation}
\begin{equation}\label{q17}
 \partial_1(R^2v_1)+\partial_2(R^2v_2) =0,
\end{equation}
where
\begin{eqnarray}\label{q14}
& H_Q(x_1,\partial_1 S,x_2,\partial_2 S) = H(x_1,\partial_1 S,x_2,\partial_2 S) &
\nonumber \\ 
& +Q_1(x_1,x_2)+Q_2(x_1,x_2) , &
\end{eqnarray}
\begin{equation}\label{q16}
 Q_a=-\frac{\hbar^2}{2m_a} \frac{\partial_a^2 R}{R} ,
\end{equation}
\begin{equation}\label{vdef}
 v_a=\frac{1}{m_a}\partial_a S ,
\end{equation}
and $H(x_1,\partial_1 S,x_2,\partial_2 S)$ is the classical Hamiltonian. 
We see that (\ref{q13}) has the same form as the classical 
time-independent Hamilton-Jacobi equation (\ref{c3}), 
with the replacement
\begin{equation}
V(x_1,x_2) \rightarrow V_Q(x_1,x_2)=V(x_1,x_2)+Q(x_1,x_2), 
\end{equation}
where $Q=Q_1+Q_2$.

In the Bohmian interpretation of QM, quantum particles have definite
trajectories analogous to classical trajectories (\ref{c4})
\begin{equation}\label{c4q}
 \frac{dX_a(t)}{dt}=\frac{1}{m_a}(\partial_a S)_t ,
\end{equation}
with the only difference that now $S$ satisfies 
the quantum Hamilton-Jacobi equation (\ref{q13}),
instead of the classical Hamilton-Jacobi equation (\ref{c3}). 
So, if $X_2(t)$ is known, then the motion of the
remaining particle can be calculated from
\begin{equation}\label{c4q0}
 \frac{dX_1(t)}{dt}=\frac{1}{m_1}(\partial_1 S_c)_t ,
\end{equation}
where $S_c(x_1,t)=S(x_1,X_2(t))$. Equivalently, it can also be calculated from
\begin{equation}\label{c4q1}
 \frac{dX_1(t)}{dt}=\frac{1}{m_1}(\partial_1 \tilde{S}_c)_t ,
\end{equation}
where $\tilde{S}_c(x_1,t)$ satisfies the quantum analogue of (\ref{c8})
\begin{equation}\label{c8q}
\tilde{H}_{Qc}(x_1,\partial_1 \tilde{S}_c,t)=-\partial_t \tilde{S}_c ,
\end{equation}
with
\begin{eqnarray}\label{c8q.1}
\tilde{H}_{Qc}(x_1,\partial_1 \tilde{S}_c,t) & = & \frac{(\partial_1 \tilde{S}_c)^2}{2m_1}
+V_c(x_1,t)
\nonumber \\
& &
+Q_{1c}(x_1,t)+Q_{2c}(x_1,t) .
\end{eqnarray}
Analogously to (\ref{c10}), we see that 
\begin{equation}\label{c10q}
 \tilde{S}_c(x_1,t)=S_c(x_1,t) +f(t) ,
\end{equation}
where $f(t)$ is some function which depends only on $t$.

Now consider (\ref{q17}) along the trajectory $x_2=X_2(t)$. Clearly,
(\ref{q17}) for $x_2=X_2(t)$ can be written as
\begin{equation}\label{q17c}
 \partial_1(R_c^2v_{1c})+(\partial_2 R^2)_c \, v_{2c} + R_c^2 (\partial_2 v_2)_c =0 .
\end{equation}
On the other hand, we have
\begin{equation}\label{evol-velocity}
 \partial_t R_c^2 = \left( \frac{\partial R^2}{\partial x_2} \right)_{x_2=X_2(t)}  \frac{dX_2(t)}{dt} 
=(\partial_2 R^2)_c \, v_{2t}.  
\end{equation}
This can be used to eliminate $(\partial_2 R^2)_c$ from (\ref{q17c}), which leads to
\begin{equation}\label{non-cons}
 \frac{v_{2c}}{v_{2t}} \, \partial_t R_c^2 + \partial_1(R_c^2v_{1c}) + R_c^2 (\partial_2 v_2)_c =0.
\end{equation}

\subsection{Conditional Schr\"{o}dinger equation}
\label{SEC3.3}

Eqs.~(\ref{c8q}), (\ref{c8q.1}), (\ref{c10q}), and (\ref{non-cons}) are exact equations.
In Appendix \ref{SEC-app} we show that they can be combined into an exact
pseudo-Schr\"{o}dinger equation. Here, however, we are more interested 
in the case when the trajectory $X_2(t)$ can be approximated by a classical trajectory.
This happens when $R(x_1,x_2)$ does not depend much on $x_2$.
(For a more detailed discussion of the classical limit in Bohmian mechanics, see also
\cite{durr-class}.) More precisely, if $L_2$ is a typical length scale within which
$R$ changes significantly by a change of $x_2$, then $X_2(t)$ behaves classically if
\begin{equation}\label{apr1}
L_2 \gg \hbar/P_2 , 
\end{equation}
where $P_2=m_2 \dot{X}_2$ is the momentum. In this approximation we can take
$\partial_2 R \simeq 0$, so (\ref{q16}) implies that 
\begin{equation}
 Q_2 \simeq 0 .
\end{equation}
Similarly, we have $\partial_2 Q_1 \simeq 0$, $\partial_2 Q_2 \simeq 0$,
implying that the quantum force $-\partial_2 (Q_1+Q_2)$ on the second particle
is negligible. Therefore the motion of the second particle is approximately governed 
by local classical forces, implying that velocity of the second particle does not depend much on the position
of the first particle at the same time, so
\begin{equation}\label{apr3}
 v_{2c}(x_1,t) \simeq v_{2t}(t) .
\end{equation}
Similarly, the quantity $(\partial_2 v_2)_c$ does not depend much on $x_1$, so we can write
\begin{equation}\label{apr4}
 (\partial_2 v_2)_c \simeq -\Gamma(t), 
\end{equation}
where $\Gamma(t)$ is a function which depends only on $t$.
Inserting all these approximations into equations (\ref{c8q}), (\ref{c8q.1}), (\ref{c10q}), and (\ref{non-cons}),
these equations reduce to a set of two independent approximate equations
\begin{equation}\label{qhjcond}
\frac{(\partial_1 S_c)^2}{2m_1}+V_c+Q_{1c}+\dot{f}(t) = -\partial_t S_c ,
\end{equation}
\begin{equation}\label{ap-non-cons}
 \partial_t R_c^2 + \partial_1(R_c^2v_{1c}) = R_c^2 \Gamma(t) ,
\end{equation}
where $\dot{f}(t)=df(t)/dt$ and 
all functions with the label $c$ are functions of $x_1$ and $t$.

We see that the left-hand side of (\ref{ap-non-cons}) looks like an equivariance equation,
but the right-hand side does not due to the $\Gamma$-term. To find a true equivariance equation
without the $\Gamma$-term, we define a new quantity
\begin{equation}\label{rho}
 \rho_c(x_1,t)=\frac{R_c^2(x_1,t)}{N(t)},
\end{equation}
where $N(t)$ will be fixed from the requirement that $\rho_c(x_1,t)$ should satisfy
an equivariance equation. Inserting (\ref{rho}) into (\ref{ap-non-cons}), we get
\begin{equation}\label{2sb}
 N(t)[ \partial_t\rho_c + \partial_1(\rho_c v_{1c})] +\rho_c [\dot{N}(t) +\Gamma(t) N(t)] =0.  
\end{equation}
The requirement of equivariance is the vanishing of the first square bracket in (\ref{2sb}) 
\begin{equation}\label{eqv}
 \partial_t\rho_c + \partial_1(\rho_c v_{1c}) =0 ,
\end{equation}
which is satisfied if the second square bracket in (\ref{2sb}) vanishes as well
\begin{equation}\label{N}
 \dot{N}(t) +\Gamma(t) N(t) =0.
\end{equation}
Eq.~(\ref{N}) is a differential equation with the solution
\begin{equation}\label{Nt}
 N(t)=N(0) e^{-\int_0^t dt \, \Gamma (t)} .
\end{equation}
Note that $\Gamma (t)$ is not a positive-definite function, so the exponent in 
(\ref{Nt}) may be either positive or negative for various values of $t$. 
On the other hand, we choose $N(0)$ to be positive, implying that
$N(t)$ is positive for all $t$.

Next we introduce the wave function
\begin{equation}\label{psi-tilde}
 \tilde{\psi}_c(x_1,t)=\tilde{R}_c(x_1,t) \, e^{i\tilde{S}_c(x_1,t)/\hbar} ,
\end{equation}
where
\begin{equation}\label{Rtilde}
 \tilde{R}_c(x_1,t)=\frac{R_c(x_1,t)}{\sqrt{N(t)}} ,
\end{equation}
and $\tilde{S}_c(x_1,t)$ is given by (\ref{c10q}). Now (\ref{rho}) can be written as 
$\rho_c =\tilde{R}^2_c$, so (\ref{eqv}) can be written as
\begin{equation}\label{eqv2}
 \partial_t\tilde{R}^2_c + \partial_1(\tilde{R}^2_c \tilde{v}_{1c}) =0 ,
\end{equation}
where, due to (\ref{vdef}) and (\ref{c10q}), 
\begin{equation}\label{vdeftilde}
\tilde{v}_{1c}=\frac{1}{m_1}\partial_1 \tilde{S}_c=\frac{1}{m_1}\partial_1 S_c .
\end{equation}
Similarly, (\ref{qhjcond}) can be written as 
\begin{equation}\label{qhjcond-tilde}
\frac{(\partial_1 \tilde{S}_c)^2}{2m_1}+V_c+\tilde{Q}_{1c}= -\partial_t \tilde{S}_c ,
\end{equation}
where, due to (\ref{q16}) and (\ref{Rtilde}),
\begin{equation}\label{q16tilde}
 \tilde{Q}_{1c}=-\frac{\hbar^2}{2m_1} \frac{\partial_1^2 \tilde{R}}{\tilde{R}}
= -\frac{\hbar^2}{2m_1} \frac{\partial_1^2 R}{R} .
\end{equation}
Eqs.~(\ref{eqv2}) and (\ref{qhjcond-tilde}), together with (\ref{psi-tilde}), are easily recognized
to be equivalent to the {\em conditional Schr\"{o}dinger equation}
\begin{equation}\label{cond_sch}
\left[ -\frac{\hbar^2}{2m_1} \partial_1^2 + V_c(x_1,t) \right] \tilde{\psi}_c(x_1,t) =
i\hbar\partial_t  \tilde{\psi}_c(x_1,t) .
\end{equation}
Eq.~(\ref{cond_sch}) is our final result, which represents a quantum analogue of the classical
Hamilton-Jacobi equation (\ref{c8}) with (\ref{c1.5}). Note, however, that (\ref{c8}) with (\ref{c1.5})
is an exact result, while (\ref{cond_sch}) is only an approximation.
Eq.~(\ref{cond_sch}) can also be obtained in a different way, by using a WKB-like expansion
in powers of $1/\sqrt{m_2}$ \cite{verresen}.
A similar result has also been obtained under more restrictive assumptions in \cite{durr-meaning}. 

In practical descriptions of open physical systems, effective Schr\"odinger equations 
of the form of (\ref{cond_sch}) are often taken as a starting point, with heuristic
justifications for the use of classical time-dependent potentials in quantum mechanics. 
(Recall that, in general, a classical time-dependent potential is defined as the conditional
potential (\ref{Vcgen}) in which
${\bf X}_{n+1}(t),\ldots,{\bf X}_N(t)$ are classical trajectories.)
Our detailed analysis above shows how such an effective description can be obtained 
from first principles, by using conditional wave functions.

\section{Physical relevance of the conditional Schr\"odinger equation}
\label{SEC-phys}

After the technical developments in Secs.~\ref{SEC-clas} and \ref{SEC-quant}, 
now we want to understand the physical relevance of the conditional Schr\"odinger equation.

\subsection{When different velocities cannot be distinguished}

In modeling quantum measurements, one often argues 
(either in Bohmian \cite{bohm,book-bohm,book-hol,book-durr} or non-Bohmian  
\cite{decoh1,decoh2} context)
that the ordinary Schr\"odinger equation itself, without particle trajectories,
provides a non-trivial time evolution $\Psi(x_1,x_2,t)$ of the form
\begin{equation}\label{phys1}
\psi(x_1)\phi_0(x_2) \stackrel{t}{\rightarrow} \sum_{b} c_b \psi_b(x_1)\phi_b(x_2) ,
\end{equation}
where $\psi(x_1)=\sum_{b} c_b \psi_b(x_1)$  is the initial wave function of the measured system,
$\phi_0(x_2)$ is the initial wave function of the measuring apparatus,
$\psi_b(x_1)$ are normalized eigenstates of the measured observable, and $\phi_b(x_2)$
are normalized macroscopically distinct pointer states well localized in the $x_2$-space. 
Then particle trajectories $X_a(t)$ with {\em any} velocities
\begin{equation}
 \frac{dX_a(t)}{dt}=u_a(X_1(t),X_2(t),t)
\end{equation}
satisfying the equivariance equation 
\begin{equation}\label{phys3}
 \partial_t |\Psi|^2 +\partial_1 (|\Psi|^2 u_1) +\partial_2 (|\Psi|^2 u_2) =0 
\end{equation}
will imply that $X_2$ will enter the channel $\phi_b(x_2)$ with probability $|c_b|^2$, 
in agreement with observations. For that purpose, the Bohmian velocity 
$u_a=v_a$ is not better than any other velocity $u_a$ satisfying (\ref{phys3}).
The localized wave packets $\phi_b(x_2,t)$ themselves move approximately classically
owing to the Ehrenfest theorem, even if the trajectory $X_2(t)$ 
is highly non-classical at lengths shorter than the width of the 
wave packet. 
So {\em if the only role of particle trajectories is to pick up some particular
channel  $\phi_b(x_2)$ in (\ref{phys1}), then observations cannot distinguish
between different velocities $u_a(x_1,x_2,t)$ satisfying (\ref{phys3}).}
Some even argue that it makes the very existence of such trajectories superfluous \cite{zeh,brown}.

\subsection{When different velocities can be distinguished}

In this paper, however, we point out that picking up some particular channel in (\ref{phys1}) 
is {\em not} the only role of particle trajectories.
In modeling time evolutions of the form (\ref{phys1}), one usually
describes it with a Schr\"odinger equation in which the Hamiltonian itself has an explicit dependence on time.
(For instance, one often uses a coupling between the measured system and the apparatus
that starts at some particular time $t_0$ and ends at another particular time $t_1$.) 
But fundamental Hamiltonians do not have an explicit dependence on time,
so such a description is at best an effective one. One might argue that
the time dependence comes from a classical time-dependent environment,
but if the macroscopic apparatus is to be ultimately described by QM as in (\ref{phys1}),
then, at the fundamental level, the environment should also be quantum.
On the other hand, if the closed system as a whole is in a state with definite total energy,
then no time evolution at all, including the time evolution in (\ref{phys1}), can be described 
by the Schr\"odinger equation alone. Some additional time-dependence is needed.

In this paper we propose that the time evolution of the wave function in (\ref{phys1})
is fundamentally a consequence of time-dependent particle trajectories. More precisely,
we propose that the time-dependent wave function $\Psi(x_1,x_2,t)$ in (\ref{phys1})
is actually a conditional wave function 
\begin{equation}
 \Psi(x_1,x_2,t)=\Psi(x_1,x_2,X_3(t)) ,
\end{equation}
where $X_3(t)$
represents trajectories of the particles constituting the environment. 
(Here by ``environment'' we mean the controllable environment degrees of freedom
which are responsible for preparations of quantum experiments. They should not be
mixed with uncontrollable environment degrees of freedom responsible for decoherence 
\cite{decoh1,decoh2}.)

It is known from experience that classical models of environment provide an approximation
that agrees well with observations. Essentially, a classical model of a time-dependent
environment can be reduced to a Hamiltonian with a time-dependent potential of the
form
\begin{equation}\label{V123}
 V(x_1,x_2,t)=V(x_1,x_2,X_3(t)) ,
\end{equation}
where $X_3(t)$ is a classical trajectory. But from the results of Sec.~\ref{SEC3.3},
we know that conditional wave function approximately satisfies a Schr\"odinger equation
with such a clasically time-dependent potential, provided that $X_3(t)$ can actually be approximated
by a classical trajectory. We have seen that such an approximation is justified
if the velocities $u_a$ are the Bohmian velocities $v_a$. 

On the other hand, with an arbitrary modification of velocities as in Eq.~(\ref{int2}), 
such an approximation does not work. (Even though it may work for some sufficiently small
modifications.)
The reason is the fact that $V(x_1,x_2,X_3(t))$ would not have
approximately classical dependence on time if $X_3(t)$ did not have
approximately classical dependence on time.
As a consequence, the conditional Schr\"odinger equation (\ref{cond_sch})
with a classical time-dependence of the potential could not be obtained.
Thus we conclude that observations prefer Bohmian velocities 
$u_a=v_a$ over other arbitrary modifications of the form of (\ref{int2}). 
With arbitrary modifications, or more precisely with velocities significantly deviating
from the Bohmian ones, it would be difficult to explain 
why quantum systems described by the time-dependent potentials of the form (\ref{V123})
with classical time-dependence $X_3(t)$ 
are in agreement with observations.

It is also useful to illustrate the meaning of ``significant'' deviations by more specific
examples. If a vector field 
${\bf f}_a({\bf x}_1,\ldots,{\bf x}_N,t)$ has the property
$\sum_a \mbox{\boldmath $\nabla$}_a {\bf f}_a =0$, then so does
the vector field ${\bf j}_a=\lambda{\bf f}_a$ with an arbitrary constant $\lambda$.
Therefore the velocity of the form
\begin{equation}\label{int2new}
{\bf u}_a={\bf v}_a +\lambda\frac{{\bf f}_a}{|\Psi|^2} 
\end{equation}
satisfies the equivariance equation (\ref{int1}) for any $\lambda$. Clearly, the Bohmian velocity 
${\bf v}_a$ cannot be experimentally distinguished from (\ref{int2new}) with a sufficiently 
small $\lambda$. In this case, the deviation of (\ref{int2new}) from ${\bf v}_a$ is not
significant. By contrast, a significant deviation of (\ref{int2new}) from ${\bf v}_a$ corresponds to the case
in which $\lambda$ is sufficiently large. Even if the quantum potential is exactly zero,
in which case ${\bf v}_a$ coincides with classical velocity,
the velocity (\ref{int2new}) will be very different from the classical velocity
if $\lambda$ is sufficiently large.

An even more specific example is provided by the case in which $|\Psi|^2$ in (\ref{int1})
does not depend on time. In this case one can choose ${\bf f}_a=|\Psi|^2 {\bf v}_a$, so
(\ref{int2new}) becomes
\begin{equation}\label{int2new1}
{\bf u}_a=(1+\lambda){\bf v}_a . 
\end{equation}
Thus the actual velocities are proportional to the Bohmian ones, but not equal to them.
An extreme example is provided by $\lambda=-1$, in which case particles do not move at all.
It should be obvious that particles which do not move cannot explain
why the Universe (with a time-independent total wave function) depends on time, 
despite the fact that (\ref{int1}) is satisfied. 

In addition, note that the condition (\ref{apr1}), needed for validity of the classical approximation,
is {\em not} satisfied for well-localized wave packets. Just the opposite, this condition
is best satisfied for very delocalized wave functions, such as a plane wave $e^{ik_2x_2}$.
When this condition is met, then the wave function does not look like a classical particle 
with a time-dependent trajectory. For that reason, the role of actual Bohmian particle trajectories 
is essential.

\subsection{Is conditional wave function essential?}

One might argue that our argument for preferring Bohmian velocities
over more general velocities can be reduced to the explanation
of the classical limit in QM, and that conditional wave functions
are a red herring which are not essential for the argument.  

To see if this is true,
consider the following question: If nature is quantum,
then why some objects seem to have classical trajectories?
The standard (non-Bohmian) answer, presented in a somewhat over-simplified form, is this:
Some objects have classical trajectories because localized wave packets have classical trajectories,
owing to the Ehrenfest theorem.
However, there is a problem with that answer.
If the wave function of the closed system has definite total energy,
then the wave function does not depend on time, and therefore
wave packets do not depend on time, and therefore
wave packets do not move at all.
So again, if wave packets do not move, then why do we see that
some objects move with classical trajectories?

The Bohmian formulation of QM provides a very clear answer. The objects which
move are not wave packets, but the particles.
But if Bohmian velocities are replaced
with more general velocities, then, in general, this motion
cannot be approximately classical. Therefore observations prefer
Bohmian velocities over more general velocities.

In this way one concludes that Bohmian velocities are preferred
without invoking conditional wave functions.
There is, however, a problem with the argument above.
This argument uses the assertion that ``the objects which move are not wave packets, but the
particles''. This assertion cannot be completely true because
there is experimental evidence that,
in many cases, wave packets do move.
But then again, how can moving wave packets be compatible with the total wave function
which does not depend on time? The Bohmian answer is:
because these time-dependent wave functions are conditional wave
functions. So this is why it is important to analyze the behavior
of conditional wave functions, and to show that they are
compatible with approximations which we already know to be
in agreement with observations.

\section{Conclusion}
\label{SEC-concl}

It is well known that QM has both deterministic aspects
(the time-evolution of the wave function) and probabilistic
aspects (the probabilistic interpretation of the wave function).
For a given deterministically-evolving wave function of a subsystem, 
our results do not change the well-established result 
that the {\em probabilistic} measurable predictions
of Bohmian mechanics do not depend on the choice of ${\bf j}_a$ in (\ref{int2}).
What our results demonstrate is that there are 
{\em deterministic} measurable predictions, regarding
the evolution of wave function itself, which are sensible
to different choices of ${\bf j}_a$.

More specifically,
in this paper we have presented a detailed derivation of the conditional 
Schr\"odinger equation satisfied by the conditional wave function, in which
the time dependence of the wave function comes from the time dependence of known particle trajectories.
If the known particle trajectories can be approximated by classical trajectories,
then the conditional Schr\"odinger equation turns out to coincide with a Schr\"odinger equation
in a classical time-dependent potential, which is known to be a good description
in agreement with observations. The approximation by classical trajectories is justified
if the particle velocities are the Bohmian velocities, but is 
not justified if the particle velocities are arbitrarily modified in a way that does not ruin 
validity of the quantum equivariance equation. We have used these results
to propose that Bohmian particle trajectories are the fundamental
source of time dependence in closed systems with definite energy, 
and to argue that observations prefer Bohmian velocities over other
velocities compatible with the quantum equivariance equation.

Our main line of reasoning can be summarized in a 
simple form as follows:
\begin{itemize}
\item[--] In practical applications of QM, one often uses an
approximation (which agrees with observations)
according to which some particles are described
by a Schrodinger equation in a classical time-dependent potential.
\item[--] Bohmian mechanics needs to explain why that approximation 
is justified.
\item[--] The conditional wave function of these particles 
approximately obeys the Schrodinger equation
in a classical time-dependent potential
only if other environment particles, which are responsible for the time-dependence
of the potential, follow approximately classical trajectories.
\item[--] Given that the total wave function of a closed system is time independent,
the environment particles can follow approximately classical trajectories
if the Bohmian velocity formula is assumed, but not if
the general velocity formula (significantly deviating from the Bohmian one) is assumed.
\item[--] Therefore, Bohmian mechanics justifies the approximation in the first item above 
if the Bohmian velocity formula is assumed,
but not if the general velocity formula (significantly deviating from the Bohmian one) is assumed.
\end{itemize}

Another, even more concise way to summarize our results, is to point out that we have found a
loophole in the usual proof that all velocities satisfying the quantum equivariance equation
give the same measurable predictions. The usual proof takes for granted that the evolution
(\ref{phys1}) does not depend on the existence of particle trajectories, which, indeed,
represents the core of the proof.
However, taking it for granted overlooks the idea that the wave function of a closed system should not depend on time.
When this overlooked idea is taken into account, one finds that the evolution (\ref{phys1}) does depend
on particle trajectories, because the wave function evolving with time is the conditional wave function.

\section*{Acknowledgments}
The author is grateful to R. Verresen for an inspiring discussion and for drawing attention
to his work \cite{verresen},
and to T. Norsen for discussions that stimulated further refinements in the manuscript. 
This work was supported by the Ministry of Science of the
Republic of Croatia under Contract No.~098-0982930-2864.

\appendix

\section{Time evolution in various interpretations of QM}
\label{SEC-interp}

Consider a closed system in a state with definite energy. 
How time evolution is possible in such a system? Unfortunately,
an interpretation-independent answer to this question does not exist.
In this Appendix we briefly review different answers provided by different interpretations.

{\it ``Copenhagen''-collapse interpretation.}
According to this interpretation proposed by von Neumann \cite{vN},
everything, including the observer, is described by the wave function. However, 
the time-evolution of the wave function is not always governed by the 
Schr\"{o}dinger equation. Instead, the act of observation is associated
with a wave-function collapse. The collapse introduces an additional
time-evolution in the system, not present in the evolution by the
Schr\"{o}dinger equation. In this interpretation the act of observation
plays a fundamental role, but the concept of observation itself is not described by physics.

{\it ``Copenhagen'' interpretation with classical macro-world.}
According to this interpretation, usually attributed to Bohr, quantum mechanics
can be applied only to the micro-world, not to the macro-world. The macro-world
is described by classical mechanics, so the time evolution in the macro-world
is not governed by a Schr\"{o}dinger equation. In a closed system a quantum
micro-subsystem interacts with a classical macro-subsystem, so that the time-dependence
of the latter induces a time dependence of the former.

{\it Modern instrumental ``Copenhagen'' interpretation.}
This is a widely-used practically oriented interpretation of QM (see, e.g., \cite{peres}),
in which QM is nothing but a tool used to predict the probabilities of measurement
outcomes for given measurement preparations. The measurement preparations
are freely chosen by experimentalists. The experimentalists themselves are not described by QM. 
The free manipulations by experimentalists introduce additional
time-dependence in the system not described by the Schr\"{o}dinger equation.
Within such an interpretation, the concept of wave function of the whole Universe does not make sense.

{\it Objective collapse.}
In this interpretation the Schr\"{o}dinger equation is modified by adding a 
stochastic term due to which the wave function collapses independently
on any observers. The best known example of such a modification
is the GRW theory \cite{GRW}.  

{\it Hidden variables.}
In this class of interpretations, the physical objects observed in experiments
are not the wave functions, but some other time-dependent variables $\lambda(t)$. 
Even if the wave function governed by the Schr\"{o}dinger equation is time-independent,
the ``hidden'' variable $\lambda(t)$ may depend on time.
The best known and most successful model of such variables is given by the
Bohmian interpretation \cite{bohm}, some aspects of which are studied in more detail 
in the present paper. 

{\it Statistical ensemble.}
According to this interpretation, the wave function is only a property of a statistical
ensemble of similarly prepared systems and tells nothing about 
properties of individual physical systems \cite{balen}. So if a wave function is
time-independent, it does not mean that individual systems do not depend on time.
This interpretation can be thought of as an agnostic variant of the 
hidden-variable interpretation, in the sense that the existence of hidden variables
is compatible (and perhaps even natural) with the statistical-ensemble interpretation,
but the statistical-ensemble interpretation refrains from saying anything more specific
about them.
 
{\it Consistent histories.}
In this interpretation \cite{griff}, the wave function is a tool to assign a probability
to a given time-dependent history of the physical system. In this sense, it is
similar to hidden-variable interpretations. However, to avoid non-localities
typically associated with normal hidden-variable theories, the consistent-histories 
interpretation replaces the classical propositional logic with a different kind of logic \cite{griff2}. 

{\it Many worlds.}
According to the many-world interpretation, the Universe as a whole is
nothing but a wave function evolving according to the Schr\"{o}dinger equation \cite{ever,MWI}.
So, if wave function of the Universe is a state with definite total energy, at first sight it seems
impossible to have any nontrivial time-dependence in the system. 
Nevertheless, a non-trivial time dependence can be introduced in a rather subtle way,
by redefining the concept of time itself \cite{page}. Even if $\psi(x_1,\ldots,x_N)$ does not depend on 
an evolution parameter $t$, some of the configuration variables $x_1,\ldots,x_N$ 
may represent readings of a physical clock, on which $\psi(x_1,\ldots,x_N)$
still depends. In such an interpretation of QM all probabilities are interpreted as
conditional probabilities, as reviewed also in \cite{kuchar,isham}. 

Of course, none of these interpretations is without difficulties. However, to avoid controversy
and keep neutrality, the difficulties will not be discussed. 
We expect that critical readers will immediately recognize some difficulties with most of these interpretations,
even without our assistance. 

\section{Pseudo-Schr\"{o}dinger equation for conditional wave function}
\label{SEC-app}

Consider a wave function $\psi(x,t)$ satisfying equation of the form
\begin{equation}\label{a1}
 \left[ -\frac{\hbar^2}{2m} \partial_x^2 + U(x,t)
+h(t) +i\hbar \frac{\Gamma(x,t)}{2} \right] \psi = i\hbar \partial_t \psi ,
\end{equation}
where $U(x,t)$, $h(t)$, $\Gamma(x,t)$, are some real, but otherwise unspecified functions.
Eq.~(\ref{a1}) looks similar to a Schr\"{o}dinger equation, but the $\Gamma$-term 
makes the operator in the square brackets non-hermitian, implying that time-evolution
of $\psi$ is not unitary. Besides, the functions $U(x,t)$, $h(t)$, and $\Gamma(x,t)$
may in fact be functionals of $\psi(x,t)$, in which case (\ref{a1}) is not a linear equation.
For those reasons, we refer to (\ref{a1}) as a {\em pseudo-Schr\"{o}dinger equation}.

Now let us write $\psi$ in the polar form
\begin{equation}\label{q12a}
 \psi(x,t)=R(x,t)e^{iS(x,t)/\hbar} .
\end{equation}
Using the identities
\begin{equation}
 -\frac{\hbar^2}{2m} \partial_x^2 \psi = \left[ \frac{(\partial_x S)^2}{2m}+Q 
-\frac{i\hbar}{2R^2} \partial_x(R^2 v) \right] \psi ,
\end{equation}
\begin{equation}
 i\hbar \partial_t \psi =\left[ \frac{i\hbar}{2R^2}(\partial_t R^2) -(\partial_t S) \right] \psi , 
\end{equation}
where
\begin{equation}
 Q=-\frac{\hbar^2}{2m}\frac{\partial_x^2 R}{R} , \;\;\;\;
v=\frac{1}{m}\partial_x S ,
\end{equation}
and multiplying (\ref{a1}) with $\psi^*$ from the left, 
one finds that the complex equation (\ref{a1}) is equivalent to a set of two real equations
\begin{equation}\label{ahj}
 \frac{(\partial_x S)^2}{2m}+U+Q+h=-\partial_t S ,
\end{equation}
\begin{equation}\label{anon-cons}
 \partial_t R^2 + \partial_x(R^2 v) = R^2 \Gamma .
\end{equation}

Next we want to show that the exact results of Sec.~\ref{SEC-polar} can be written in the form of
(\ref{ahj}) and (\ref{anon-cons}).
For that purpose we first observe that (\ref{c8q}), (\ref{c8q.1}), and (\ref{c10q})
can be written as
\begin{equation}\label{ahjc}
\frac{(\partial_1 S_c)^2}{2m_1} +V_c+Q_{2c}+Q_{1c}+\dot{f} = -\partial_t S_c .
\end{equation}
Second, we write (\ref{non-cons}) as
\begin{equation}\label{non-cons-a}
\partial_t R_c^2 + \partial_1(R_c^2v_{1c}) = R_c^2 \left[ 
\left( \frac{v_{2c}}{v_{2t}} -1 \right) \frac{\partial_t R_c^2}{R_c^2} - (\partial_2 v_2)_c \right] .
\end{equation}
We see that (\ref{ahjc}) and (\ref{non-cons-a}) have the same form as (\ref{ahj}) and (\ref{anon-cons}),
respectively, with the identifications $m=m_1$, $x=x_1$, $R=R_c$, $v=v_{1c}$, $Q=Q_{1c}$, 
provided that we take 
\begin{equation}\label{aU}
U(x_1,t)=V_c(x_1,t)+Q_{2c}(x_1,t), 
\end{equation}
\begin{equation}
h(t)=\dot{f}(t) ,
\end{equation}
\begin{equation}\label{aGamma}
 \Gamma(x_1,t)=\left( \frac{v_{2c}(x_1,t)}{v_{2t}(t)} -1 \right) 
\frac{\partial_t R_c^2(x_1,t)}{R_c^2(x_1,t)} - (\partial_2 v_2)_c(x_1,t) .
\end{equation}
In other words, the conditional wave function $\psi_c(x_1,t)$ satisfies the
pseudo-Schr\"{o}dinger equation (\ref{a1})
\begin{equation}\label{a1.1}
\left[ -\frac{\hbar^2}{2m_1} \partial_1^2 + U(x_1,t)+\dot{f}(t) +i\hbar \frac{\Gamma(x_1,t)}{2} \right] 
\psi_c = i\hbar \partial_t \psi_c ,
\end{equation}
where $U(x_1,t)$ and $\Gamma(x_1,t)$ are given by (\ref{aU}) and (\ref{aGamma}),
respectively. We see that (\ref{aU}) and (\ref{aGamma}) are functionals of 
$\psi_c(x_1,t)$, so (\ref{a1.1}) is a non-linear equation.

Finally, the $\dot{f}$-term in (\ref{a1.1}) can be eliminated 
by defining a new wave function
\begin{equation}
 \tilde{\psi}_c(x_1,t)=\psi_c(x_1,t) \, e^{i f(t)/\hbar} .
\end{equation}
The new wave function satisfies
\begin{equation}\label{a1.2}
\left[ -\frac{\hbar^2}{2m_1} \partial_1^2 + U(x_1,t) +i\hbar \frac{\Gamma(x_1,t)}{2} \right] 
\tilde{\psi}_c = i\hbar \partial_t \tilde{\psi}_c .
\end{equation}
A similar pseudo-Schr\"{o}dinger equation has also been obtained in \cite{oriols} and \cite{norsen}.

\end{document}